A Cryogenic SiGe Low Noise Amplifier Optimized for Phased Array Feeds.

Wavley M. Groves III and Matthew A. Morgan

NRAO Central Development Laboratory

1180 Boxwood Estate Road

Charlottesville, VA 22903-4608

A Cryogenic SiGe Low Noise Amplifier Optimized for Phased Array Feeds.

Wavley M. Groves III and Matthew A. Morgan

Abstract— The growing number of phased array feeds (PAF) being built for radio astronomy demonstrates an increasing need for low noise amplifiers (LNA) that are designed for repeatability, low noise, and ease of manufacture. Specific design features which help to achieve these goals include the use of unpackaged transistors (for cryogenic operation), single-polarity biasing, straight plug-in RF interfaces to facilitate installation and re-work, and the use of off-the shelf components.  The focal L-band array for the Green Bank Telescope (FLAG) is a cooperative effort by Brigham Young University (BYU) and the National Radio Astronomy Observatory (NRAO) using warm dipole antennae and cryogenic Silicon Germanium Heterojunction Bipolar Transistor (SiGe HBT) LNAs.  These LNAs have an in band gain average of 38 dB and 4.85 Kelvin average noise temperature.  Although the FLAG instrument was the driving instrument behind this development, most of the key features of the design and the advantages they offer apply broadly to other array feeds, including independent-beam and phased, and for many antenna types such as horn, dipole, Vivaldi, connected-bowtie, etc.  This paper will focus on the unique requirements array feeds have for low noise amplifiers and how amplifier manufacturing can accommodate these needs.

I.  INTRODUCTION

The next generation of the National Radio Astronomy (NRAO) and Brigham Young University (BYU) collaboration of the focal L-band array for the Green Bank Telescope (FLAG) phased array feed required a new cryogenic LNA (low noise amplifier) to be developed in order to obtain optimal performance and replace the previous amplifiers [1] with ones using superior performing SiGe HBT devices.  Phased Array Feeds (PAFs) are multi-beam focal-plane arrays for radio telescopes.  In contrast to conventional horn arrays, the PAF antenna elements are usually smaller, more closely spaced, and over-illuminate the dish aperture.  Multiple independent beams are formed by electronically combining the signals from multiple antenna elements.  In this way, the formed beams are independently steerable within the primary field-of-view, can be closely spaced or even overlapping on the sky, and the beam pattern may be shaped to improve aperture efficiency or create a null in the direction of a strong interferer.  PAF projects are becoming more common [1]-[14] so it was decided that these LNAs should be developed to fill the needs of PAF projects as they evolve and the number of antennae grows making it easier to scale and fill these needs in shorter timelines with easily produced and reliable amplifiers that exhibit repeatable results.  These LNAs are competitive with those of the two current cryogenic PAFs: the 19-element, dual-polarized, fully cryogenic PAF prototype camera for the Arecibo radio telescope (AO19 cryo-PAF) and National Research Council Canada (NRC) cryoPAF4, falling between the respective quoted noise figures of 14 Kelvin and 3.5 Kelvin.  The performance quoted here is near state-of-the-art for LNAs in this frequency range [15] and combined with the manufacturability and ease-of-assembly afforded by the features discussed in this paper, the FLAG instrument is expected to have a superior figure of merit (system temperature over aperture efficiency) than any other astronomical phased array currently available.

II.     REQUIREMENTS FOR CRYOGENIC LOW NOISE AMPLIFIERS

The new generation of cryogenic FLAG LNAs inherited a few requirements from the previous attempt of the phased array feed.  Consisting of nineteen dual-polarized dipoles, the array needs nineteen dual-channel LNAs for thirty-eight LNA signal paths, forty if you count the spare module.  The new array incorporated these requirements of the cryogenic LNAs to maintain consistency with key elements of the PAF.  We refined the LNA package in order to suit the specialized needs of phased array feeds, facilitating ease of integration into future projects.  It is worth noting that the array-specific requirements called out in sub-sections A-F below apply not just to the FLAG instrument using dipole antennas, but equally well to other PAF antenna element technologies, such as Rocket, Vivaldi, Chequerboard, etc.

A.      Unpackaged transistors.

The use of bare die unpackaged transistors is necessary because the original FLAG LNAs demonstrated a high mechanical failure rate of plastic packaged devices when thermally cycled between cryogenic and ambient temperatures.  This failure is due to mechanical strain resulting from mismatched thermal-expansion coefficients.  Bare die transistors are not susceptible to the parasitic effects of packaging a device and therefore behave more predictably when cryogenically cooled.  When schedule permits, experimentation with packaging such as ceramic needs to be explored in NRAO's cryogenic receivers, this time it was necessary to be safe and use unpackaged devices.

B.      Simple, single polarity bias.

The bias needs for thirty-eight cryogenic LNA channels could be a significant thermal load if the amplifiers were servo biased due to the amount of wiring needed to control that gate and drain of a HEMT device, so the bias requirements needed to be as simple as possible. Using a single-polarity bias supply mitigates the issues of servo biasing while meeting the requirements of the PAF.  Limited space in the cryogenic dewar requires the bias connector to be compact and easy to insert.

C.      Spacing of input ports.

The placement of the dipoles dictates the spacing of the input ports on the LNA chassis. The LNAs are connected to a vacuum feedthrough comprising a custom quartz bead and center conductor. The cold side of the plate that holds the dipoles has soldered coaxial stainless steel thermal transitions permanently attached. This spacing can be adjusted with minimal effort to accommodate different antennae configurations.

D.      Mechanically designed for ease of installation.

Space is at a premium in the FLAG cryogenic dewar and the LNAs need to be easily installed and removed by technicians for maintenance, ease of wiring, and ease of heat strapping. The original FLAG LNA, permanently soldered to the end of these transitions, made the LNA unwieldly to work on once completely assembled making repairs difficult, this issue can be mitigated by using push on connectors.

E.      Easily produced in large quantities.

Large numbers of LNAs are needed for the growing size of PAFs so they should be able to be quickly produced using as many common off the shelf components as possible to reduce component lead time and cost.

F.      Reliability and repeatability.

The LNAs need to be impedance-matched, reliable, and repeatable to ensure optimum performance of the array. These LNA modules are required to be consistently and easily manufactured to eliminate troubleshooting poorly performing modules due to assembly error or unreliability of design as much as possible.

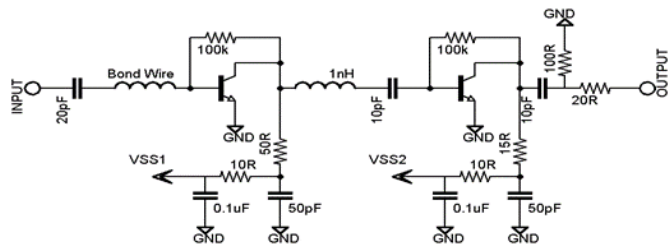

Fig. 1.  A simplified EAGLE schematic of a PAF cryogenic LNA channel.

III.    ELECTRONIC DESIGN

Based on an earlier prototype SiGe HBT S-band (1.7-2.6 GHz) [15] amplifier model and scaled for L-band (1.2-1.7 GHz) using simulation software, an inductor was added to the circuit to low pass the amplifier for a more fine-tuned frequency response. This stable and satisfactory model using software tuning and simulation tools provided a schematic for use in a PCB layout program to create the dual amplifier module (Fig. 1). The FLAG antenna elements were optimized such that the beam-dependent active impedance presented to the amplifiers would be nearly optimal over the FOV for amplifiers with a standard 50 Ohm noise- and gain-match [1]. The amplifiers were therefore matched for optimum noise measure at 50 Ohm input impedance, based on available information this is the most common impedance for multi-feed instrumentation.

The board layout took great care to ensure that ground pours and small chassis walls accommodated by slots in the circuit board isolated the high gain elements.  The bias traces were kept isolated as possible by distance and ground pour complimented by a number of vias added to tie the top and bottom ground planes together both electrically and thermally.  The use of a printed circuit board (PCB) also offered an advantage in assembly time over chip and wire amplifiers.  Chip and wire amplifiers require the accurate placement of components installed by hand to ensure symmetry in layout and consistency of the LNA.  The RF portion of the PCB for the two amplifiers in the package is perfectly symmetrical requiring only the attachment of components without the need to measure out their placement.  The circuit board layout employed long traces for the feedback resistors to ensure optimal placement when tuning for performance.

A Rogers 4350b [16] substrate was chosen for the thermal and mechanical properties it exhibited during the development of the earlier S-band prototype amplifier in which the author has experimented with multiple substrate materials.  The surface finish of Electroless Nickel / Electroless Palladium / Immersion Gold (ENEPIG) was chosen for the ability to bond with both gold and aluminum bonding wire as well as solder [17].  The ENEPIG surface was important because the tuning bond wire needed to be gold, which is the standard process that NRAO uses for cryogenic LNAs; the SiGe HBT devices have aluminum bonding pads to accommodate industry standard aluminum wire.  Concerns over possible cryogenic series resistance of the aluminum wire due to its silicon doping kept these bonds short as possible.  Gold wire used for connection to the devices would have fallen victim to purple plague, an aluminum-gold intermetallic compound [18], and caused the device to be unusable.

A customer-defined transistor layout fabricated by ST Microelectronics, comprising a 5-finger HBT with 15um emitter length, exhibits excellent noise parameters when cooled to 15 Kelvin [19]-[22]. These devices are not packaged and therefore not susceptible to the mechanical damage caused to previous thermoplastic packaged transistors when thermally cycled.  The bipolar SiGe HBT devices only require a single polarity power supply, thus keeping the bias wiring for the nineteen LNA modules simple and the least possible thermal load.  The aforementioned attributes made these particular SiGe HBT devices an excellent choice for the FLAG project.

Standard off the shelf passive components were chosen to facilitate ease of manufacture and to eliminate the need for custom or rare components with long lead times for ordering.

IV. CHASSIS DESIGN

Careful consideration was required in respect to the mechanical design of the LNA chassis.  The previous LNA for the FLAG had input ports soldered directly to the LNA chassis to mitigate the loss introduced by a coaxial connector.  This limited the size of the amplifier chassis to dimensions that could fit through the port in the cryogenic dewar wall and it was permanently attached to the plate which held the dipole assembly. This size restriction compromised the PCB layout and limited connector choices.

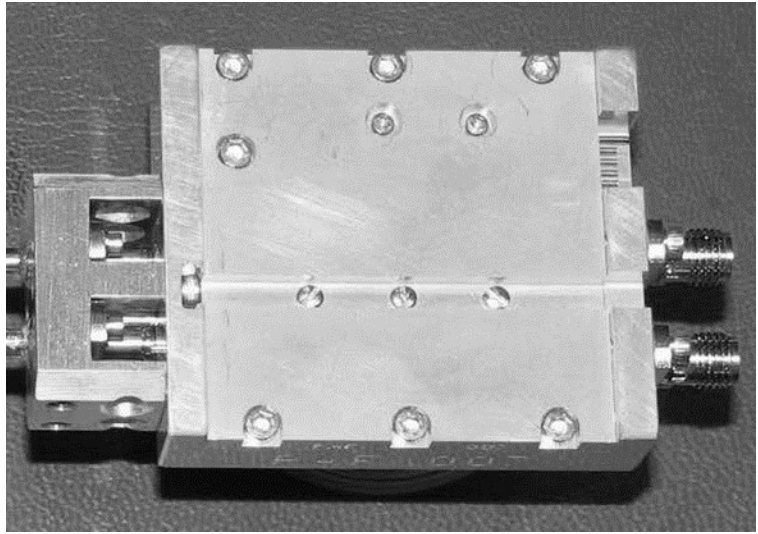

Fig. 2.  A dual LNA package and mating block showing the slots machined for access with a driver.

To resolve the limitations of the original chassis design, the new LNA employed push on SMP connectors to enable the amplifiers to blind mate to a block on the end of the thermal transitions (Fig. 2).  In order to make sure the weight of the LNA was not too much of a strain on the mechanical connection of the SMP connectors, captive screws used to connect firmly to the LNA to the mating block/dipole assembly (Fig. 3) designed at the Green Bank Observatory (GBO).  This ensured that the push on connections limited the compromising effects of gravity or the stress of thermal cycling.  Two slots were machined into the top and bottom of the chassis of the LNA to accommodate a driver to tighten the captive screws and a positive stop machined into the mating block at the engage length of the SMP bullet to protect the bullet from being crushed by over tightening the screws or by tightening them unevenly.  Low profile screw heads used in these slots minimized interference with the driver.  The spacing of the ports is too narrow to accommodate the barrels of SMA connectors on the input of the dual LNA package, the proximity of other LNA packages also made using a wrench on the input unwieldly, and the use of SMP connectors eliminated the need to access the connector from the side with an open-ended wrench.

The new LNA's output connectors were SMA to maintain consistency with the cabling of the previous generation and were spaced to accommodate a wrench to tighten the connections.  An Omnetics strip connector was used to provide the bias voltages because of the small size and low cost, but lead times on a custom part make a Nano-D connector a better choice for subsequent revisions for off the shelf availability and mechanical stability.

The chassis employs two walls to isolate the high gain sections of the amplifiers to reduce the chances of feedback loops and to isolate the bias traces as much as possible (Fig. 4) and slots cut into the PCB to accommodate these walls.  To keep the bias traces as far from each other as possible the bias lines for the channel one amplifier were routed on the bottom of the PCB and around the areas machined into the chassis to provide maximum thermal transfer and ground contact between the chassis and PCB.  Multiple screws firmly attach the PCB to the chassis for a good mechanical and thermal connection.

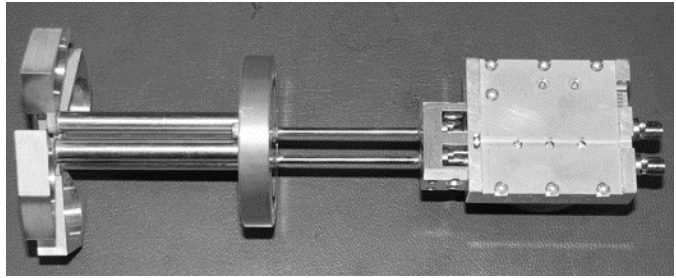

Fig. 3. A dual LNA package assembled with dipole antennae, dewar window, coaxial thermal transitions, and mating block.

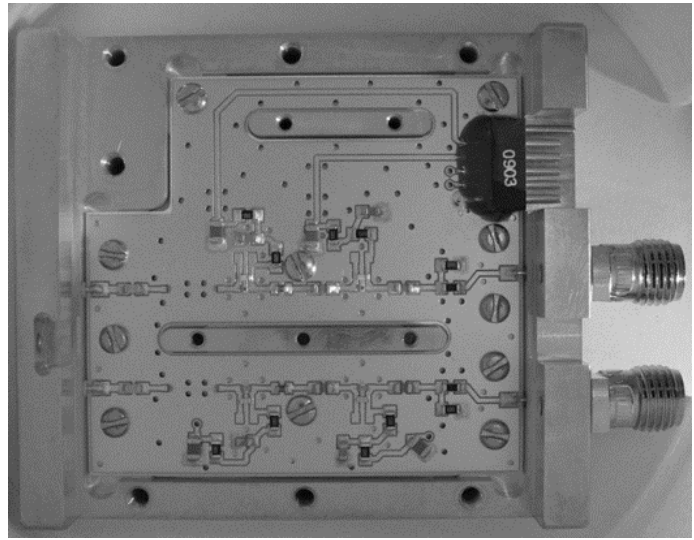

Fig. 4. An assembled dual LNA package with the cover removed showing the PCB and component layout.

The prototypes and twenty LNA production run were done on un-masked PCBs with all components placed by hand and attached with EPO-TEK H20e conductive epoxy for maximum flexibility in reworking the circuit thus eliminating the need to solder and/or clean flux near unpackaged transistors, minimizing risk of damage to these sensitive components. In the future, solder mask will be used on the PCBs to prevent potential solder bridges and all commercial components placed and soldered in a fabrication house so that only unpackaged transistors and feedback resistors need placement by hand. The feedback resistors are placed by hand after wire bonding because their proximity to the transistors interfere with the bonding tool when placing the transistor ground bonds. This minimizes the time and effort skilled technicians need to build LNAs allowing larger scale production numbers and lower person-hours needed to put large numbers of LNAs in arrays by only focusing their efforts on the crucial elements of the circuit.

V. TEST RESULTS

The first prototype had similar measured and modeled noise characteristics but with about 10 dB more gain than modeled due to the ground bond lengths on the transistors and placement of the feedback resistors. With shortened bond wires and the feedback resistors moved closer to the device, the second prototype's results were consistent with the simulated model and production dual LNA packages.

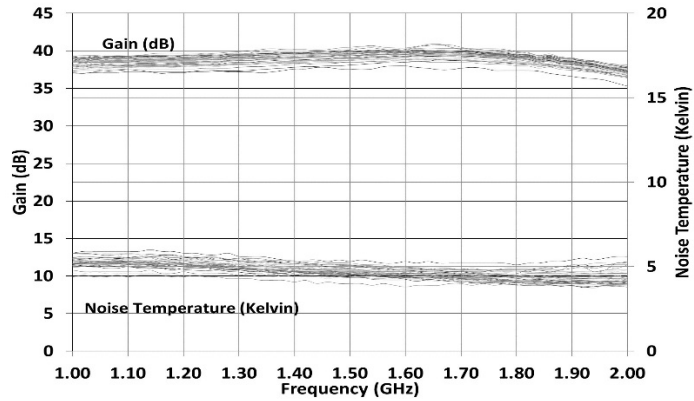

Fig. 5. Forty channels of LNA test data compiled measured at 15 Kelvin.

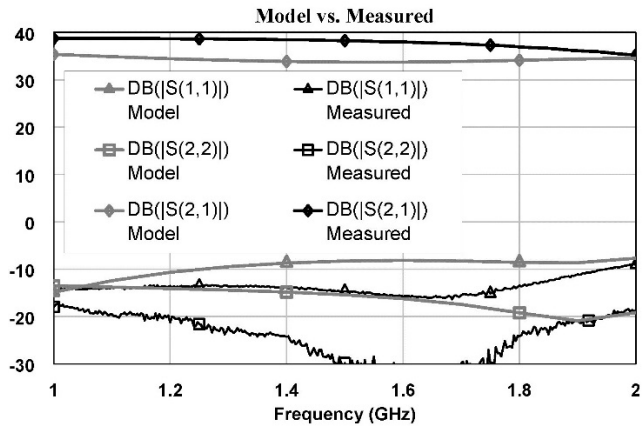

Fig. 6. The warm signal data of a typical LNA (black) vs. model (grey).

A few small changes were made to the PCB and chassis after prototyping for ease of assembly and reliability before building the final forty (nineteen dual channel packages and one spare dual channel package) LNA channels.  These included extra screws firmly fixing the PCB against the chassis near the transistors eliminating the need for the PCB attachment with H20e epoxy.  A second small chassis wall and corresponding slot in the PCB was added to further isolate the bias traces for the first stage and to allow for an RF tight seal.

In the target bandwidth of 1.2 GHz to 1.7 GHz the LNAs exhibited gain averaging around 38 dB and noise average of 4.85 Kelvin when cooled to 15 Kelvin in a cryogenic test set. Fig. 5 shows the consistency of both gain and noise temperature of all forty channels of the production LNAs.  This is at least a few Kelvin improvement over the previous generation of amplifiers used in prior PAF builds at the observatory, and is in advantage to the FLAG project, which aims to have a lower noise temperature than any other astronomical phased array currently demonstrated.

The warm signal data was measured by an Agilent E8364B PNA network analyzer (Fig. 6). The typical LNA exhibits good gain, input return loss, and output return loss within the target frequency range of 1.2 GHz to 1.7 GHz and the results are very repeatable with all of the LNA channels built.

The cold data for the same typical LNA channel when measured in a cryogenic dewar at 15 Kelvin has excellent gain and noise parameters as acquired by an Agilent N8975A Noise Figure Analyzer using the cold attenuator method of measurement [23], [24]. This method combines a warm noise diode outside the cryogenic Dewar with a cooled attenuator inside to achieve an Excess Noise Ratio (ENR) suitable for LNA's operating at these temperatures. The uncertainty of these measurements is thought to be about ± 1K, limited primarily by calibration of the Excess Noise Ratio (ENR) of the noise diode. This is the standard method of measurement for LNAs with coaxial RF connectors at NRAO and yields repeatable data when measuring cryogenic LNA modules at 15 Kelvin, the expected operating temperature of the LNAs on the telescope. The resulting LNA had surprisingly flat broadband gain and noise from 250 MHz to 1.9 GHz (Fig. 7) making the amplifier suitable for a broadband application or alternate frequency range with minimal need for optimization.

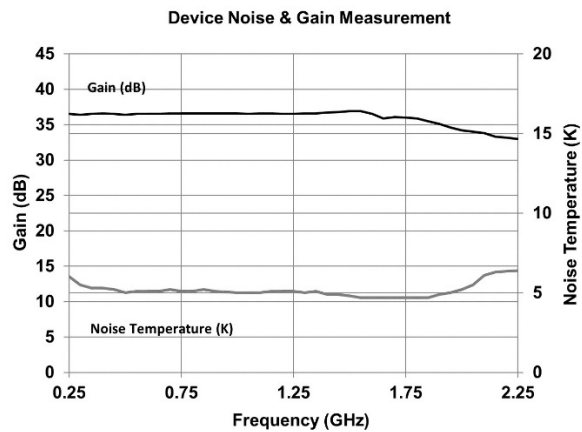

Fig. 7. Broadband cryogenic test data of the second prototype LNA.

VI. CONCLUSION

Cryogenic SiGe HBT low-noise amplifiers were designed and manufactured to exhibit a consistent 4.85K average noise temperature and 38 dB of gain. This average noise temperature will enable the FLAG array to achieve world-record sensitivity for astronomical PAFs. The LNAs were cryogenically tested and found to exhibit highly repeatable performance over twenty dual-channel units. Design decisions such as: unpackaged SiGe HBT transistors, single polarity bias, spacing of input ports to accommodate connection to antennae, mechanically designed for ease of installation, easily produced in large quantities, and reliability and repeatability. These design decisions were made to solve challenges specifically related to phased-array feed and multi-beam instrumentation.

ACKNOWLEDGMENT

This research was carried out at the National Radio Astronomy Observatory (NRAO), Charlottesville, VA. NRAO is a facility of the National Science Foundation (NSF) operated under cooperative agreement by Associated Universities Inc.  Dipole/LNA mating block designed by Bob Simon of the GBO.  The SiGe HBT devices, manufactured by ST Microelectronics and acquired under sub-contract by Dr. Sander Weinreb of Caltech

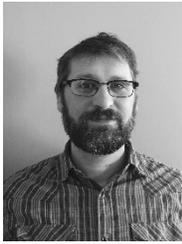

**Wavley M. Groves III** received the A.A.S.E.T. degree from the Piedmont Community College, Charlottesville, in 2011.  From 2005 to 2007, he was a Professional Audio Repair Technician at Diversified Audio, Inc. in Tampa, FL and manufactured avionics and ultrasound medical test equipment for UMA, Inc. in Dayton, VA before joining the Low Noise Amplifier Group at the National Radio Astronomy Observatory (NRAO) as a Senior Technician in 2007.  Mr. Groves is currently a Technical Specialist at NRAO and member of the CDL's Integrated Receiver Development Program which he joined in 2012 where he is involved in the design, prototyping, and troubleshooting of prototype instrumentation refining it to become reliable radio astronomy solutions and is the Recording Engineer and Head Technician at EccoHollow Art & Sound.

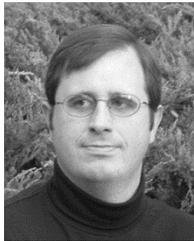

**Matthew A. Morgan** received the B.S.E.E. degree from the University of Virginia, Charlottesville, in 1999, and the M.S. and Ph.D. degrees in electrical engineering from the California Institute of Technology, Pasadena, in 2001 and 2003, respectively.  From 1999 to 2003, he was an Affiliate of the Jet Propulsion Laboratory (JPL), Pasadena, CA, where he developed monolithic millimeter-wave integrated circuits (MMICs) and multi-chip modules (MCMs) for atmospheric radiometers and spacecraft telecommunication systems.  Dr. Morgan is currently a Scientist / Research Engineer with the National Radio Astronomy Observatory (NRAO), Charlottesville, VA, where he is involved in the design and development of low-noise receivers, components, and novel concepts for radio astronomy instrumentation in the cm-wave, mm-wave, and submm-wave frequency ranges. He was Project Engineer for the K-Band Focal Plane Array on the Green Bank Telescope, and is currently the head of CDL's Integrated Receiver Development Program.